\input harvmac
\input epsf

\font\ticp=cmcsc10
 
\def\Title#1#2{\rightline{#1}\ifx\answ\bigans\nopagenumbers\pageno0\vskip1in
\else\pageno1\vskip.8in\fi \centerline{\titlefont #2}\vskip .5in}

\font\ticp=cmcsc10

\def\vp{\varphi}
\def\hr{{\hat{\rho}}}
\def\hit{{\hat{t}}}
\def\hphi{{\hat{\varphi}}}
\def\p{\partial}
\def\ssc{\scriptscriptstyle}
\def\ie{{\it i.e.,}\ }
\def\eg{{\it e.g.,}\ }

\def\hz{{\hat{z}}}
\def\l#1{\ell_{#1}}
\def\la{\lambda}

\lref\rCvetic{For a review of domain wall solutions in supergravity, see
M. Cvetic,  Phys. Rept. {\bf 282} (1997) 159, hep-th/9604090.}
\lref\rWald{R. Wald, {\sl General Relativity}, University of Chicago Press,
Chicago (1984) p. 179.}
\lref\rGT{J.~Garriga and T.~Tanaka,{\it ``Gravity in the brane-world,''}
hep-th/9911055.}
\lref\rGKR{S. Giddings, E. Katz, and L. Randall, in preparation.}
\lref\rBTZ{M.~Ba{\~n}ados, C.~Teitelboim and J.~Zanelli,
Phys.\ Rev.\ Lett.\ {\bf 69} (1992) 1849,
hep-th/9204099.}
\lref\rCG{A.~Chamblin and G.W.~Gibbons, {\it ``Supergravity on the
Brane,''} hep-th/9909130.}
\lref\rKW{W.~Kinnersley and M.~Walker, Phys.\ Rev.\ {\bf D2}
(1970) 1359.}
\lref\rPD{J.F.~Plebanski and M.~Demianski, Ann.\ Phys.\ {\bf 98} (1976)
98.}
\lref\rEJM{R.~Emparan, C.V.~Johnson and R.C.~Myers,
Phys.\ Rev.\ {\bf D60} (1999) 104001, hep-th/9903238.}
\lref\rCHR{A.~Chamblin, S.W.~Hawking and H.S.~Reall,
{\it ``Brane-world black holes,''} hep-th/9909205.}
\lref\rMann{R.B. Mann, Class.\ Quant.\ Grav.\ {\bf 14} (1997)  L109,
gr-qc/9607071.}
\lref\rRS{L.~Randall and R.~Sundrum,
{\it ``An alternative to compactification,''} hep-th/9906064.}
\lref\rRSfirst{L.~Randall and R.~Sundrum,
Phys.\ Rev.\ Lett.\ {\bf 83} (1999) 3370, 
hep-ph/9905221.}
\lref\rautre{other people who invented Randall and Sundrum?
See e.g., M.~Gogberashvili,
{\it ``Hierarchy problem in the shell-universe model,''}
hep-ph/9812296. 
For a review of earlier work and references, see 
J.M.~Overduin and P.S.~Wesson,
Phys.\ Rept.\ {\bf 283} (1997) 
gr-qc/9805018.}
\lref\rtalk{R.C. Myers, {\it ``On the Gravity of the AdS/CFT
Correspondence,''} talk at  ITP Conference on Strong Gravitational Fields,
June 22-26, 1999,  http://www.itp.ucsb.edu/online/gravity\_c99/myers/.}
\lref\rpositive{G.T.~Horowitz and R.C.~Myers,
Phys.\ Rev.\ {\bf D59} (1999) 026005, hep-th/9808079.}
\lref\rIsrael{W. Israel, Nuovo Cim. {\bf 44B} (1966) 1.}
\lref\rMTW{C.W.~Misner, K.S.~Thorne and J.A.~Wheeler,
{\it Gravitation} (San Francisco: Freeman, 1973).}
\lref\rNemanja{N.~Kaloper,
{\it ``Bent domain walls as braneworlds,''} hep-th/9905210;
T.~Nihei, {\it ``Inflation in the five-dimensional universe with an
orbifold extra  dimension,''}
hep-ph/9905487; H. B. Kim and H. D. Kim, {\it ``Inflation and Gauge Hierarchy
in Randall-Sundrum Compactification"}, hep-th/9909053.}
\lref\rHS{G.T.~Horowitz and H.J.~Sheinblatt,
Phys.\ Rev.\ {\bf D55} (1997) 650, gr-qc/9607027.}
\lref\rstand{G.W.~Gibbons and S.W.~Hawking,
Phys.\ Rev.\ {\bf D15} (1977) 2752.}
\lref\rthree{A.~Ashtekar and M.~Varadarajan,
Phys.\ Rev.\ {\bf D50} (1994) 4944, gr-qc/9406040.}
\lref\rthreeo{S.~Deser, R.~Jackiw and G.~'t Hooft,
Ann.\ Phys.\ {\bf 152} (1984) 220.} 
\lref\rGL{R.~Gregory and R.~Laflamme,
Phys.\ Rev.\ Lett.\ {\bf 70} (1993) 2837,
hep-th/9301052.}
\lref\rEHM{R. Emparan, G. Horowitz and R. Myers, {\it
``Exact Description of Black Holes on Branes,"} hep-th/9911043.}
\lref\rMP{R.C.~Myers and M.J.~Perry,
Ann.\ Phys.\ {\bf 172} (1986) 304.}
\lref\rGubser{S.S.~Gubser,
{\it``AdS/CFT and gravity,''}
hep-th/9912001, and references therein.}

\baselineskip 16pt
\Title{\vbox{\baselineskip12pt
\line{\hfil hep-th/9912135}}}
{\vbox{
{\centerline{Exact Description of Black Holes on Branes II:}}
\vskip 1ex
{\centerline{Comparison with BTZ Black Holes and Black Strings}}
}}
\centerline{\ticp Roberto Emparan$^a$, Gary T. Horowitz$^b$, 
Robert C. Myers$^c$}
\vskip 2ex
\centerline{\it $^a$ Departamento de F{\'\i}sica Te\'orica,
Universidad del Pa{\'\i}s Vasco, Apdo.\ 644, E-48080 Bilbao, Spain}
\vskip 1ex
\centerline{\it $^b$ Physics Department, University of
California, Santa Barbara, CA 93106 USA}
\vskip 1ex
\centerline{\it $^c$ Institute for Theoretical Physics, University of
California, Santa Barbara, CA 93106 USA}
\vskip 1ex
\centerline{\it $^c$ Department of Physics, McGill University, Montr\'eal,
QC, H3A 2T8, Canada}
\vskip 2ex
\centerline{$^a$wtpemgar@lg.ehu.es, $^b$gary@cosmic.physics.ucsb.edu,
$^c$rcm@hep.physics.mcgill.ca}

\vskip 2cm
\centerline{\bf Abstract}
\bigskip
We extend our recent discussion of four-dimensional black holes bound to a 
two-brane to include a negative cosmological constant 
on the brane. We find that for large masses, the
solutions are precisely BTZ black holes on the brane, and BTZ `black strings'
in the bulk. For smaller masses, there are localized black holes which
look like BTZ with corrections that fall off exponentially.
We compute when the maximum entropy configuration changes
from the black string to the black hole.
We also present exact solutions describing rotating black holes
on two-branes which are either asymptotically flat or asymptotically $AdS_3$.
The mass and angular momentum on the brane agree with that in the bulk.

\Date{December, 1999}

\newsec{Introduction}

In a recent paper \rEHM\ we constructed exact solutions describing four
dimensional black holes bound to a two-brane in anti-de Sitter
(AdS) space. This construction was motivated by the
suggestion \rRS\ that lower dimensional gravity is naturally induced on
a brane (domain wall) placed in AdS spacetime \rCvetic. The original
discussion involved linearized perturbations of a three-brane in
AdS$_5$, but there were questions raised \rCG\ about the boundary
conditions used near the AdS horizon. By considering a two-brane in
AdS$_4$, we were able to investigate the viability of this scenario by
studying the properties of exact black hole solutions.

We found that lower dimensional gravity was indeed reproduced on the
brane at large distances, with no difficulty arising at the AdS horizon.
(The horizon geometry was, in fact, unchanged.) Large black holes
appeared as flattened pancakes with a much smaller extent off the brane
than along the brane. These features will extend to the five
dimensional case. However, there were some awkward features of our
solutions. The effects of four-dimensional gravity were significant even
at scales much larger than the bulk cosmological constant. This is
perhaps not surprising given that $2+1$ gravity (without a cosmological
constant on the brane) does not have any black hole solutions. Another
unusual feature of $2+1$ gravity is that there is a maximum total mass.
We found that as the mass approached this limit, the size of the black
hole grew without bound. Thus one can have arbitrarily large black holes
with finite total mass. 

In this paper we extend our previous analysis to allow a negative
cosmological constant in the $2+1$ gravity theory induced on the brane.
This has the advantage that both of the unusual features of pure $2+1$
gravity mentioned above are avoided. When a negative cosmological
constant is added, $2+1$ gravity does have black hole solutions \rBTZ,
and there is no upper limit on the total mass. Thus one might hope that
this is a more realistic model of the higher dimensional physics.

The solutions in \rEHM\ were obtained by starting with a metric
describing an accelerating black hole in AdS$_4$. We then introduced the
two-brane by cutting the spacetime along an appropriate surface and
gluing it to a copy of itself. To add a cosmological constant on the
brane we start with a slightly more general solution, but in this case
we have to compactify the fourth direction by introducing two different
branes. We will argue that this is a general feature of branes with negative
curvature and is not special to $AdS_4$. 
Unlike previous discussions of compactifications with two branes
\rRSfirst, both of our branes have positive tension.

We find that large
black holes on the brane agree exactly  with the $2+1$ dimensional BTZ
solution. 
However, these black holes are not localized near the brane. They extend
across the fourth dimension and form a `BTZ black string'.
It is perhaps not surprising that when the size of the 
extra dimension is finite,
large black holes on the brane correspond to black strings in the bulk.
This is what one
expects from a standard Kaluza-Klein compactification.
For smaller masses, we have two different black hole
solutions in which the horizon does not reach the second brane. In both
cases, the solution looks like the BTZ black hole with correction terms
that fall off exponentially in the proper distance.
These terms are negligible outside the horizon
for the solution with smaller area.
 Off the brane, it looks like 
the BTZ black string  with its ends capped off. In other
words, in this case, the horizon looks more like a cigar than a pancake.
The solution with larger horizon area can have significant deviation from
the BTZ solution outside the black hole.

Another motivation for considering a negative cosmological constant on the
brane is the following.
It has been
argued in higher dimensions that black strings are unstable
when their width shrinks to less than the AdS scale \rCHR.
Since we now have exact solutions for both the black strings and
localized black holes, we can examine this transition in
detail. We find that indeed the maximum entropy configuration 
changes from the black string to the black hole when the minimum
cross-sectional area is of order the AdS radius. 

For black holes on asymptotically flat two-branes, it was shown \rEHM\ that
the mass measured asymptotically on the brane agreed with a 
four dimensional thermodynamic mass obtained by integrating the relation
$dM=TdS$ in the bulk. We show that the same is true for black holes
on asymptotically $AdS_3$ branes.

We also construct solutions describing rotating black holes on a
two-brane both with and without a cosmological constant on the brane. In
the first case, we recover on the brane the metric of rotating BTZ black
holes with some modifications. In the latter case, the metric on the
brane is locally equivalent to the metric on the equatorial plane of the
four-dimensional Kerr solution. We will see that the rotation of the
black hole in the four-dimensional space is transferred to the brane in
such a way that the angular momentum detected on the brane is precisely
equal to the four-dimensional angular momentum.

\newsec{The AdS C-metric and two-branes}

We begin with the following solution to Einstein's equation with negative
cosmological constant describing accelerating black holes in AdS$_4$ \rPD:
\eqn\cads{ds^2={1\over A^2 (x-y)^2} \left[ H(y)dt^2 -{dy^2\over H(y)} 
+{dx^2\over G(x)} +G(x) d\varphi^2 \right],}
where
\eqn\hy{H(y)=-\la +ky^2 -2 mAy^3}
\eqn\gx{G(x)=1+ kx^2 -2 mAx^3}
and $k=+1,0,-1$. The metric satisfies $R_{\ssc AB}=-(3/\l4^2)
g_{\ssc AB}$,  where
\eqn\defell{\l4 = {1\over A\sqrt{\la +1}} }
sets the scale for the bulk cosmological constant
\ie $\Lambda_4=-3/\l4^2$. Note that from eq.~\defell, we require
$\la\ge-1$. In the limit $\la \rightarrow -1$ (with $k=-1$),
we recover the C-metric \rKW\ describing a pair of black holes accelerating
in an asymptotically flat spacetime. The general solution with $\la > -1$ is
called the AdS C-metric. This describes black holes accelerating in AdS$_4$. 
The parameters $m$ and $A$ are related to the mass
and acceleration of the black hole (at least for certain
limits --- see below).

Another interesting limit reduces the metric \cads\ to that
of static four-dimensional black holes in AdS$_4$ \rMann.
This limit turns off the acceleration parameter $A$ as follows:
set $y=-1/(rA)$, $t= A\bar t$, $\la=-\Lambda_4/3A^2>0$, and
take the limit $A\rightarrow0$ while keeping $\bar t, r, x$, $\varphi$ 
and $\Lambda_4$ finite. In this way $m$
gives the mass parameter of the static black hole solutions
of Einstein's equations with a negative cosmological
constant $\Lambda_4$. Recall that
the horizons of the static $k=-1$ black holes are topological spheres and
have finite area. In contrast, the $k=0,+1$ black holes have
horizons with the topology of $R^2$ and infinite area,

The  special case of \cads\ with $\la =0$ and $k=-1$
was employed in \rEHM. Many of the descriptive comments made there
still apply for the general metric \cads. For example, due to the
overall factor $(x-y)^{-2}$, the solution does not include points with
$x=y$. Generically, these points correspond to the boundary of
the asymptotically AdS geometry.
Hence, we will restrict $y$ to always be less than $x$.
There is also a curvature singularity at $y=-\infty$.
Clearly both $t$ and $\vp$ are Killing coordinates.
As the notation indicates $\vp$ is an angular coordinate and each zero of
$G(x)$ corresponds to an axis for the rotation symmetry. 
Similarly $t$ may be regarded as the time coordinate, and
each zero of $H(y)$ corresponds to a Killing horizon for $\partial_t$.
The smallest zero $y_0$ of $H(y)$ defines the black hole event horizon.
Typically one finds that the accelerating black holes have
horizons with the topology of $R^2$ and infinite area, for all
$k$ values.\foot{Note that this infinite area is not a problem
for the present analysis. As in \rEHM, the infinite tail of the horizon
will be cut out by the introduction of a two-brane and we will 
obtain a black hole of finite size.} Only for $k=-1$ and $mA$ not `too
large', is the event horizon a sphere with  finite area.

To orient ourselves with the metric \cads, let us set $m=0$. 
The geometry then has constant curvature and is locally the same as
AdS$_4$. To cast the metric into a more familiar form
define
\eqn\changek{r={\sqrt{y^2+\lambda x^2}\over 
A(x-y)},\qquad
\rho=\sqrt{1+kx^2\over y^2+\lambda x^2}.}
Then \cads\ becomes
\eqn\nice{ds^2={dr^2\over{r^2\over \l4^2}-\lambda}+r^2\left[ -(
\lambda\rho^2 -k)dt^2+{d\rho^2\over 
\lambda\rho^2-k}+\rho^2d\varphi^2\right].}
These forms of the AdS$_4$ metric had been considered earlier in
\rEJM. The metric in the brackets
has constant Riemann curvature ${\cal R}^{\mu\nu}{}_{\rho\sigma} = 
-\lambda \left(\delta^\mu{}_\rho \delta^\nu{}_\sigma - \delta^\mu{}_\sigma
\delta^\nu{}_\rho\right)$. When $m=0$,  $\lambda$ can be rescaled by rescaling
the coordinates $r$ and $\rho$, and so one may set $\la=0$ or $\pm1$.
These three choices correspond to three distinct ways of slicing AdS$_4$.
Notice that $\lambda =0,\ k=-1$ corresponds to the usual Poincare invariant
slices.

To introduce a brane into the spacetime, we need a surface whose
extrinsic curvature is proportional to the intrinsic metric. Then we can
cut the spacetime off at the surface, take two copies of one side, and
glue them together. The resulting space will have a delta-function stress
tensor along the surface which is proportional to the induced metric,
\ie\ a two-brane.
The extrinsic curvature of a surface can be computed via
\eqn\extrin{
K_{\mu\nu}=\nabla_{\!\mu} \,n_{\nu}
={1\over2}n^\sigma\p_\sigma  g_{\mu\nu}}
where $\mu,\nu$ run over the components tangent to the surface, and
$n^\sigma$ is the unit outward pointing normal vector. (The last
formula results in the special case that $\partial_\mu n^\sigma=0$,
which will apply in the cases of interest for this paper.) For example,
a short calculation reveals that slices at constant $r$ 
in eq.~\nice\ satisfy $K_{\mu\nu}=r^{-1}\sqrt{r^2/\l4^2-\lambda}
\;g_{\mu\nu}$. Hence the space can be cut at any of these surfaces and
glued to a copy of itself to construct domain walls (two-branes).

Since a surface  $r=L$ has constant curvature, there is a 
cosmological constant $\Lambda_3=-\lambda/L^2$ on the brane. Hence
there are three distinct cases: {\it (i)} $\la=0$, in which case the
radial slices are flat. Here we may only choose $k=\pm1$. The choice
$k=-1$ yields global Minkowski coordinates with $t$ timelike everywhere.
{\it (ii)} $\la<0$, in which case the radial slices have the geometry
of three-dimensional de Sitter space, $dS_3$. With $k=-1$,
$t$ is timelike for small $\rho$ and there are de Sitter horizons
on the two-brane at $\rho=|\lambda|^{-1/2}$. The latter correspond to
$y^2=-\lambda$ in the C-metric coordinates of eq.~\cads. 
{\it (iii)} $\la>0$, where locally the constant $r$ slices have the
geometry of AdS$_3$. The three possible values of $k$ yield
different parameterizations of AdS$_3$: $k=-1$ corresponds to
global coordinates, whereas $k=0$ and $k=+1$ yield the metrics for
massless and massive BTZ black holes \rBTZ\ (after $\vp$ is periodically
identified --- see next section). This case with $\la>0$ will be
the primary focus of our investigations below.

In their constructions, Randall and Sundrum introduced branes on 
Poincare invariant slices of AdS, \eg $r$ = constant in eq.~\nice\ 
with $\la=0$, $k=-1$. There were two distinct scenarios. Originally
they considered compactifying the transverse dimension by the introduction
of two branes on either end of a finite interval in AdS \rRSfirst.
In this case, one finds that the inner brane must have a negative tension.
However, in a second scenario,
they also found lower dimensional gravity arose for
a single brane and an infinite transverse dimension \rRS. 
These discussions were extended to de Sitter slicings
to give a description of inflationary cosmology within the
context of the Randall-Sundrum constructions \rNemanja. In this case,
one may again consider two distinct scenarios. Either one has two branes
bounding a finite interval, one with positive tension and the other
with negative tension, or one has a single brane. In the latter case,
however, the transverse direction is not infinite since the brane
is spherical and encompasses a finite region at the
center of AdS. Following the discussion above, these constructions
would correspond to placing the branes along surfaces of constant $r$
with $\la<0$ in
eq.~\nice. 

Finally one may consider the Randall-Sundrum construction
for branes with negative curvature. A subtlety arises in this
case which we will illustrate explicitly in four dimensions, but the same
comments also apply for five or higher dimensions.
In eq.~\nice\ with $\la>0$, $g_{rr}$ is
singular at $r=\sqrt{\la}\l4$ but this is simply a coordinate singularity
\rEJM. The latter is most easily seen by making the coordinate
transformation $r = \sqrt\lambda\l4 \cosh u$, with which
eq.~\nice\ becomes
\eqn\nicer{ds^2={\l4^2du^2}+\la\l4^2\cosh^2u\left[ -(
\lambda\rho^2 -k)dt^2+{d\rho^2\over 
\lambda\rho^2-k}+\rho^2d\varphi^2\right].}
Here we see the entire spacetime is covered by $-\infty < u < \infty$.
However there are now two asymptotic regions where the scale factor
for the negative curvature slices grows without bound, \ie $u\rightarrow
\pm\infty$. Consider introducing a single brane by cutting
along a surface of constant  $u$, removing
the region with large positive $u$ and gluing an identical copy of
the remaining geometry.  The resulting space would still have
two asymptotic regions as $u\rightarrow-\infty$ on either side of the
single brane. In this situation, the `zero mode' bound
to this brane would not be normalizable. Alternatively, one
might say that it would require an infinite amount of energy to excite
this mode \rEHM. In any event, one would not find an effective
lower dimensional theory of gravity on the brane. Hence for the negative
curvature slicings, one must introduce two branes by making two
cuts and removing both asymptotic regions at $u\rightarrow\pm\infty$.
Gluing this geometry to an identical copy of itself produces a space
where the gravitational `zero mode' would be normalizable and one would find
gravity is effectively lower dimensional on large distance
scales.

Another interesting feature arises here because with the negative
curvature slicing, the AdS space effectively contains a `throat.'
That is, the scale factor multiplying the metric on each slice, \ie
$r^2=\la\l4^2\cosh^2u$, decreases monotonically
to a minimum value at $r=\sqrt{\la}\l4$
(or $u=0$) as we approach this surface from either of the asymptotic regions.
Hence if we cut the metric \nicer\ along
one surface at constant positive $u$ and the other at constant negative $u$,
and remove both asymptotic regions at large $|u|$, then the extrinsic curvature
on both ends will be a positive multiple of the metric. Hence both
of the branes in this construction will have a positive tension,
in contrast to the cases with $\la\le0$, or with $\la>0$ when both
cuts are made in the positive $u$ region.\foot{The `throat' also
allows for other exotic constructions. For example,
one could produce a single brane by cutting at $u=\pm U_0$ and
identifying these two surfaces.}

For a Randall-Sundrum construction with two negative curvature branes,
we now compute the relationship between the four-dimensional Newton's constant
and that for the effective three-dimensional theory of gravity on the branes.
The metrics under consideration are of the form
\eqn\metr{ds^2={dr^2\over{r^2\over \l4^2}-\lambda}+{ r^2\over L_1^2}\,
g_{\mu\nu}(x)\,dx^\mu dx^\nu\ .}
We are assuming that one of the branes lies on the slice at $r=L_1$
(on the positive $u$ patch),
so $g_{\mu\nu}$ is precisely the induced metric on this brane.
This metric will satisfy ${\cal R}_{\mu\nu}=-(2\la/L_1^2)\,g_{\mu\nu}$.
We will assume that the second brane lies at $r=L_2$ on the patch
with $u<0$ (and so the induced metric there is
$(L_2/L_1)^2g_{\mu\nu}$). Now as in standard Kaluza-Klein compactification,
we have to integrate over the volume between the two branes. Hence we
integrate $r$ from $L_1$ to the minimum radius $\sqrt \lambda\l4$
and then back out to $L_2$ (on the negative $u$ patch).
Since our interest is in relating $G_3$ to $G_{4}$, we
focus on the Einstein-Hilbert term in the action. The brane tensions will
be tuned in order to produce the required cosmological constant
on the brane. The reduced effective action is then
\eqn\effcosm{\eqalign{
I'=&{2\over16\pi G_4}\left(\int_{\sqrt{\lambda}\l4}^{L_1}
+\int_{\sqrt{\lambda}\l4}^{L_2}\right) dr
{r/L_1\over \sqrt{{r^2\over
\l4^2}-\lambda}}\,\int\!d^3x\,\sqrt{-g}\,{\cal R}(g)+\ldots
\cr
=&{1\over16\pi G_{3}}\int\!d^3 x\,\sqrt{-g}\,\left( {\cal
R}(g)+{2\lambda\over L_1^2}\right)\cr}}
The radial integral is easily evaluated and we obtain
\eqn\gneut{2{\l4\over L_1}\left(\sqrt{L_1^2-\la\l4^2}+
\sqrt{L_2^2-\la\l4^2}\right)G_3=G_4\ .}
Note that we have inserted the appropriate cosmological constant
term in the three-dimensional effective action, where the three-dimensional
AdS scale is $\l3=L_1/\sqrt{\la}$.
We emphasize that this calculation is normalized for physicists
living on the brane at $r=L_1$.

Now we turn to applying the Randall-Sundrum construction to
the general solution \cads\ with $m>0$.
It is clear from the $m=0$ discussion that we have to introduce two
different branes. In fact,
the asymptotic region far from the black hole corresponds to $x,y \rightarrow
0$. It follows from \hy\ and \gx\ that in this limit, the effects of the
black hole become negligible and the spacetime approaches  \nice.
As usual, one can introduce a two-brane on surfaces where the extrinsic
curvature is proportional to the metric. So we need to find two such surfaces
in the AdS C-metric \cads. For a surface of constant $x$,
this will be true when $G'(x)=0$. The simplest choice (and the one used in 
\rEHM) is to take the surface $x=0$. 
Similarly, one can use a surface of constant $y$ provided that $H'(y)=0$.
The simplest choice is again $y=0$. So to construct our solution we
take the points $(x,y)$ with $x\ge 0$ and $y\le 0$ in \cads\ and glue 
this region to an identical
copy along the boundaries $x=0$ and $y=0$.\foot{Other possible locations
for the branes will be considered briefly in the discussion section.}
 It is natural to foliate the
quadrant $x\ge 0, \ y\le 0$ with surfaces of constant slope $x/y$. Since
$r$ is a function only of $x/y$ \changek, these are surfaces of constant 
$r$.\foot{Even though $r$ was originally defined for $m=0$, it is convenient
to introduce it even for $m>0$. This is because the solution far from the
black hole always approaches the AdS$_4$ metric \nice.}
The surface $x=0$ corresponds to  $r=1/A$ and $y=0$  corresponds to 
$r=\sqrt\lambda/A$. However, note that these surfaces
are on opposite sides of the `throat'. This can be seen in eq.~\changek\ 
where as $x/y$ decreases from zero to minus infinity,
$r$ decreases from $1/A$ to its minimum value, and then increases
back to $\sqrt \lambda/A$. So these surfaces both have positive extrinsic
curvature (this can also be checked directly)
and both correspond to positive tension branes. Notice that the brane at $x=0$
cuts through the black hole horizon, and contains (for $m\neq 0$) the
singularity at $y=-\infty$. In contrast, the brane at $y=0$ is
completely non-singular and does not contain any black hole horizons.
These choices for the two branes correspond to $L_1=1/A$ and
$L_2=\sqrt\lambda/A$ in the previous paragraph. Using eq.~\defell,
a bit of algebra then reduces eq.~\gneut\ to:
\eqn\gnewt{G_3={A G_4\over 2}\ .}
Notice that this is independent of $\lambda$ and curiously, it agrees with
the relation derived with a single brane and $\lambda=0$, \ie\ $A=1/\ell_4$
\rEHM. Further with the above choice of $L_1$, the effective
three-dimensional AdS scale is given by
\eqn\ascale{\l3^2={1\over A^2\la}\ .}
Note that $\l3>\l4$, where the four-dimensional AdS scale is given in
eq.~\defell. It will be useful to keep in mind that the parameters
$A$ and $\la$ together
determine the cosmological constant on the brane and the bulk,
while $m$ is related to the (localized) black hole on the brane, which we
discuss below.

\newsec{Black holes on the brane.}

The AdS C-metric with $\lambda =0$ was discussed extensively in \rEHM.
Here we examine the case $\lambda > 0$.

\subsec{BTZ black strings}

As noted above, when $m=0$ the metric reduces to eq.~\nice\ and if $k=+1$,
surfaces of constant $r$ have the geometry of BTZ black holes \rBTZ\ when
we simply make a periodic identification of $\vp$ with some period $\Delta\vp$.
The horizon is at $\rho = 1/\sqrt\la$.
The entire spacetime describes a `BTZ black string' in AdS$_4$.  When the
branes are introduced as above, the black string extends between $r=L_1=1/A$
on the positive $u$ patch and $r=L_2=\sqrt{\lambda}/A$ on the negative
$u$ patch. These black strings
are analogous to those constructed in \rCHR, except that we are considering
a construction with two branes.

Including the contributions from the two AdS geometries
that have been glued together, the total area of the event horizon is
\eqn\btzstring{{\cal A}={2\Delta\varphi \over A^2\sqrt{\lambda}}.} 
Note that at $r=L_1$, the horizon has a proper circumference
of ${\cal C}_1=\Delta\vp/A\sqrt{\la}$. The circumference shrinks
to ${\cal C}_{min}=\Delta\vp/A\sqrt{\la+1}$ at $r=r_{min}=\sqrt{\la}\l4$,
and then expands again to ${\cal C}_2=\Delta\vp/A$
as the string extends to $r=L_2$.

To determine the effective mass, we compare the induced metric at
$r=L_1$ with the standard BTZ metric \rBTZ\foot{In the literature on BTZ
black holes, it is common to set $G_3=1/8$.}
\eqn\btzmet{
ds^2=-\left({\hr^2\over\l3^2}-8G_3M_3\right)d\hit^2
+\left({\hr^2\over\l3^2}-8G_3M_3\right)^{-1}d\hr^2+\hr^2d\hphi^2}
where $\hphi$ has periodicity $2\pi$, and $\ell_3$ is the three-dimensional
radius of curvature \ascale.
To produce this form, we must rescale the coordinates in eq.~\nice\ 
as
\eqn\rescale{
t={A\Delta\vp\over2\pi}\hit\ ,
\qquad
\rho={2\pi A\over\Delta\vp}\hr\ ,
\qquad
\vp={\Delta\vp\over2\pi}\hphi\ .}
A short calculation then shows that  the total three-dimensional mass is 
\eqn\massstr{ M_3={1\over 8G_3}\left({\Delta\varphi\over 2\pi}\right)^2.}
Note that the metric \btzmet\ has a conical-like singularity at
$\hr=0$. This singularity then appears on each constant $r$ slice of the
BTZ black string and so extends throughout the four-dimensional spacetime.

\subsec{BTZ-like black holes}

We are, however, more interested in solutions with localized four-dimensional
black holes on the brane, so we turn to the case with $m>0$. 
The case $m=0$ and $k=1$ considered above was special in that
$G(x)$ never vanished,
so $\vp$ could have any periodicity.
For $m>0$, this periodicity is fixed by demanding that the
geometry is smooth on the axis where $G(x)$ vanishes.
With $m>0$, $G(x)$ always has one (and only one) positive root,
$x_2$, and a conical singularity on 
this axis is avoided by setting
$\Delta\varphi$ to 
\eqn\delphi{\Delta\varphi ={4\pi \over |G'(x_2)|}.}

Recall that in our construction of a black hole localized on a
brane, the portion of the space with $x<0$ is cut out and so
$x$ is restricted to lie in the range $0\le x\le x_2$.
Now if we let $\rho=-1/y$ the geometry induced on the
two-brane at $x=0$ is 
\eqn\wallgeo{ds^2={1\over
A^2}\left[-\left(\lambda\rho^2 -k-{2mA
\over\rho}\right)dt^2+ \left(\lambda\rho^2 -k-{2mA\over\rho}\right)^{-1}
d\rho^2 +\rho^2 d\varphi^2\right].}
This is similar to the locally AdS metrics, but with extra terms of 
the form $2mA/\rho$
coming from the four-dimensional nature of the black hole.
While these corrections have power law  decay in $\rho$, they are
actually decaying exponentially in the radial proper distance which is
asymptotically $ R_{\rm proper}=\l3\ln \rho$.
This should be expected since we have compactified
the transverse direction by introducing two branes. Hence, we should
find a discrete spectrum of massive Kaluza-Klein modes for the fluctuations of
the four-dimensional metric \rRSfirst. This behavior can be contrasted
to that for asymptotically flat branes with an infinite transverse
dimension. There one finds power law corrections to the metric \rEHM,
or to the Newtonian potential \refs{\rRS,\rGT,\rGKR}, 
arising from a continuum of
massive `Kaluza-Klein' modes.

As $\rho\rightarrow\infty$ we recover the geometry of the
surface $r=1/A$ in \nice. 
By rescaling the coordinates as in eq.~\rescale, we can compare this
asymptotic geometry to the standard BTZ form \btzmet\ in order
to determine the three-dimensional mass. The result of this
calculation is
\eqn\massads{ M_3={k\over 8G_3}\left({\Delta\varphi\over 2\pi}\right)^2.}
Using \delphi\  and \gx, it is useful to re-express this mass as
\eqn\mthreex{M_3={k\over 2G_3}{x_2^2\over
(3 +k\,x_2^2)^2}.}
Now the horizon of the four-dimensional black hole 
is at the negative root of $H(y)$,
$y=y_0$. It extends off the brane at $x=0$ to the maximum value  of $x$, $x_2$.
Using \changek, the latter corresponds to 
the radial coordinate
\eqn\outtor{r_0={\sqrt{y_0^2+\lambda x_2^2}\over A(x_2-y_0)}.}
The area of the horizon is finite and equal to
\eqn\fntarea{{\cal A}= {2\Delta \vp\over A^2}\int_0^{x_2} {dx\over (x-y_0)^2}
= {2\Delta\varphi\over A^2}{x_2\over |y_0|(x_2+|y_0|)}.}
Finally note that the brane metric \wallgeo\  contains a
curvature singularity at $\rho=0$. This singularity, however, is
confined to the brane, and is not present on
test two-branes at $r\ne 1/A$, \ie in the induced metric on constant
$r$ slices for $r\ne 1/A$. 

Now recall that $k$ can take the values 0 or $\pm 1$. So we discuss
each of these individual cases in turn below:

\noindent ({\it i}) {$k=+1$:}

This is the most interesting case since with $k=1$, eq.~\wallgeo\ is
similar to the BTZ metric, but with extra  $mA/\rho$ corrections.
If $2mA \ll \lambda^{-1/2}$, these extra terms are negligible outside the
horizon, and so the exterior geometry is essentially identical
to that of a BTZ black hole. However, for $2mA\gg\lambda^{-1/2}$,
there will be significant deviations from the
BTZ metric outside the black hole.

Surprisingly, we see from eq.~\mthreex\ that 
$M_3$ vanishes both when $mA\rightarrow 0$ ($x_2\rightarrow \infty$) and
when $mA\rightarrow \infty$ ($x_2\rightarrow 0$). This 
implies that, in contrast to BTZ strings,
there is a maximum  mass for the  black holes that are localized to
the brane. The maximum mass occurs at
$mA=2/\sqrt{27}$ ($x_2=\sqrt{3}$), where
\eqn\maxmthree{M_{3,{\rm max}}={1\over 24 G_3}.}
This is a very small  mass of the order the three-dimensional
Planck mass. The effect is entirely due to
four-dimensional dynamics. It would appear that one cannot localize a
large mass black hole on the brane. The only solution for large mass would
the BTZ black string. Even though $M_3$ vanishes when $mA\rightarrow 0$,
it is not proportional to $m$ even in this limit. 
For small $mA$, $x_2 \approx 1/(2mA)$, 
and so we have
\eqn\mthree{G_3 M_3=2 (mA)^2.}

There are two limits in which we can easily examine the geometry
of the black hole horizon: small $mA$, $2mA\ll {\rm min}\lbrace 1,\la^{-1/2}
\rbrace$, and large $mA$, $2mA\gg {\rm max}\lbrace 1,\la^{-1/2}\rbrace$.

For small $mA$, $x_2\approx 1/2mA$ and $y_0\approx -\sqrt\la$. 
In this limit, the area of the event horizon \fntarea\ 
approaches that of the black string with the same mass (same $\Delta
\vp$). This is less surprising if one observes that as
$mA$ goes to zero, $x_2$ goes to infinity, and $r_0$ approaches the
position of the second brane at $r=\sqrt\la/A$. That is, the horizon
nearly stretches across the bulk AdS space to the second brane.
With a closer examination, one realizes that
outside the horizon, $y_0 \le y <0$, the function $H(y)$
is essentially unchanged by a small $mA$. Furthermore,
$G(x)$ is unchanged until
$x$ becomes of order $1/mA$, at which point it rapidly goes to zero.
So the geometry near these `localized' black holes is very similar
to that of the BTZ black string off the brane. In fact, 
a small `localized' black hole is really a black string
which is capped off at the ends!\foot{These black holes would then
realize the `cigar-shaped' horizon geometries envisioned in ref.~\rCHR.}
One way of understanding the fact that small BTZ-like black holes 
cannot be localized near the brane very well is just that there are no small
black holes in AdS$_4$ which look like BTZ on a slice (see the case
$k=-1$ below).  

For large $mA$, $x_2\approx (2mA)^{-1/3}$ and $y_0\approx -(\la/2mA)^{1/3}$.
Above, it was noted that the mass measured on the brane goes to zero
in this limit.
However, the black hole area tends to a finite value
\eqn\constarea{{\cal A}={8\pi\over 3 A^2}{1\over
\lambda^{1/3}(1+\lambda^{1/3})}.}
Similarly, the circumference of the horizon on the brane, ${\cal C}\rightarrow
4\pi/3A\lambda^{1/3}$. Further in this limit, $r_0\approx \la^{1/3}/A\sqrt{
1+\la^{1/3}}$
and the proper distance that the black hole extends off the brane
is finite. Note that all of these quantities depend only on
$\lambda$, but not on $m$. The interpretation of the
result for $r_0$ depends on the value of $\lambda$.
First note that in general $r_{min}=\sqrt{\la}\l4$ is achieved 
in eq.~\changek\ at $x=|y|/\la$ with negative $y$.
Hence given the values of $x_2$ and $y_0$ above, we see that for
small $\la$, $r_0$ actually sits on the positive $u$ side of the
`throat' and so the event horizon has a pancake geometry.
As $\la$ increases, the black hole becomes fatter and 
touches the `throat' at $\la=1$.
For large $\lambda$, the horizon extends through the
`throat' onto the negative $u$ patch.

Of course, the most surprising effect found in the large $mA$ regime
is that one can find
black holes of finite size with vanishingly small mass on the brane.
This is reminiscent of the fact that for asymptotically flat branes,
$\la=0$, one can construct arbitrarily large black holes with finite
energy \rEHM. This seemed to be related to the fact that there is a
maximum total mass  in $2+1$ gravity without a cosmological constant.
We now see that even though there is no a priori maximum mass when
the cosmological constant is negative, there is still an upper limit on 
the mass of a localized black hole on the brane. This results in finite
black holes with arbitrarily small mass.

In the section 4.1, we will examine the stability of these BTZ-like
black holes as compared to the BTZ strings, on the basis of their
relative entropies.

\noindent ({\it ii}) {$k=-1$:}

With $k=-1$, the metric on the brane \wallgeo\ is then identical to a
slice through the equator of the (spherically symmetric)
Schwarzschild AdS metric. The situation
is now similar to the case of an asymptotically flat brane \rEHM: When
$mA$ is small, the horizon is approximately spherical, but when $mA$ is
large, the black hole is a flattened pancake. Hence in these geometries,
the black hole is very effectively confined to the vicinity of the
brane. However, one crucial
difference is that the black holes with $k=-1$ have {\it negative} mass
on the brane, as seen in eq.~\massads. Note that the mass of these
black holes varies monotonically from $M_3=0$ for $x_2=0$ ($mA\rightarrow
\infty$) to $M_3=-1/8G_3$ for $x_2=1$ ($mA\rightarrow0$).
Precisely this range of masses arise in the negative mass solutions
in AdS$_3$ constructed in \rBTZ. This sector interpolates between
the global ground state of AdS$_3$ (with mass $M_3=-1/8G_3$) and the
massless BTZ black hole. In pure $2+1$ gravity these solutions have
(naked) conical singularities. In the present context, these singularities
can be hidden behind the horizons of four-dimensional black holes.

\noindent ({\it iii}) {$k=0$:}

The $k=0$ solutions are very similar to the case of large $mA$
and $k=+1$ (or $k=-1$). This is not new: it is
known that in AdS space, the horizons of static spherical or
hyperbolic black holes become flat in the limit of infinite mass (this
is sometimes called the ``infinite volume limit,'' particularly in the
context of the AdS/CFT duality). In the present situation, this can be
achieved by scaling the mass and the coordinates as $(x,y,t) \to
\alpha^{-1/3} (x,y,t)$, $m\to \alpha m$, and letting $\alpha\to\infty$.

\newsec{Black hole thermodynamics}

We would now like to consider the thermodynamic properties of these
BTZ-like black holes on AdS$_3$ branes. Consider first the Hawking
temperature. As usual we can define the Hawking temperature as
$T=\kappa/2\pi$ where $\kappa$ is the surface gravity of the black
hole. There is a small difficulty since $\kappa$ depends on the
normalization of the timelike Killing vector
$\xi\propto\p_t$. In asymptotically flat spacetimes, one usually requires
that $\xi$ have unit norm at infinity, but it is not obvious what the
proper normalization is in the present case where the branes
are asymptotically AdS.
To resolve this difficulty, we
can turn to the thermodynamics of the BTZ black hole \btzmet. The
circumference of the
horizon is ${\cal C}=2\pi r_{\ssc H}=4\pi\l3\sqrt{2G_3M_3}.$ Hence the
entropy satisfies
\eqn\btzfirst{
S=\pi\l3\sqrt{2M_3\over G_3}\quad\rightarrow\quad
{dS\over dM_3}={\pi\l3\over\sqrt{2G_3M_3}}={1\over T}\ .}
Now one finds that this temperature matches precisely the surface
gravity of the Killing vector $\xi=\p_t$, or alternatively $\beta=1/T$
is the periodicity of the euclidean coordinate $\tau=i t$. Hence this
suggests that if we normalize the coordinates such that
the asymptotic metric on the AdS$_3$ brane is
\eqn\btzasym{
ds^2\simeq-{\hr^2\over\l3^2}d\hat t^2
+{\l3^2\over \hr^2}d\hr^2+\hr^2d\hat \vp^2}
where $\hat \vp$ has periodicity $2\pi$, then the Hawking temperature will
be the surface gravity associated with the Killing vector $\p_t$ in this
coordinate system. So having decided on the appropriate normalization,
a short calculation shows that 
\eqn\btzT{
T={A\over2\pi}{|H'(y_0)|\over |G'(x_2)|}={A\over2\pi}{|y_0|\over x_2}
{3mA|y_0|+1\over 3mAx_2-1}\ .}
The black hole entropy is simply determined by the
four-dimensional area \fntarea
\eqn\btzentropy{
S={{\cal A}\over4G_4} = 
{\Delta \vp \over 2G_4A^2}{x_2\over|y_0|(x_2+|y_0|)}.}

\subsec{Equality of three and four dimensional masses}

For the case of asymptotically flat branes, it was shown \rEHM\
that the three-dimensional mass measured asymptotically on the brane
agreed precisely with  a four-dimensional thermodynamic
mass obtained by integrating
the first law 
\eqn\firstb{
\delta M=T\,\delta S.}
We now show that this is also true when there is a negative cosmological
constant on the brane. It is impractical to try to express $T$ in terms
of $S$ in order to compute $M$. Instead, it proves more convenient to
express both $S$ and $T$ in terms of a single auxiliary variable. An
appropriate choice is
\eqn\btzz{
\hz=|y_0|/x_2}
for which one has various identities such as
\eqn\biden{\eqalign{
x_2&={1\over2mA}{\lambda+\hz^2\over\lambda-\hz^3}\ ,\cr
|y_0|&={\hz\over2mA}{\lambda+\hz^2\over\lambda-\hz^3}\ ,\cr
2mA&={(\lambda+\hz^2)\hz\sqrt{1+\hz}\over(\lambda-\hz^3)^{3/2}}\ .\cr}}
Note that $\hz$ is a monotonic function of $mA$, and that
\eqn\bidenm{\eqalign{
mA\rightarrow 0\ ,\qquad&\hz\simeq 2mA\sqrt{\lambda}\cr
mA\rightarrow \infty\ ,\qquad&\hz\simeq\lambda^{1/3}\ .\cr}}
One also has that $x_2$ and $|y_0|$ are monotonic
functions of $\hz$ in the range $0\le\hz\le\lambda^{1/3}$.

Now in terms of $\hz$, the Hawking temperature \btzT\ becomes
\eqn\btzTT{
T={A\hz\over2\pi}{2\lambda+3\lambda\hz+\hz^3\over
\lambda+3\hz^2+2\hz^3}\ .}
Let us also consider
\eqn\btzDT{
{\p T\over\p\hz}={A\over\pi}{(\lambda-\hz^3)
(\lambda+3\lambda\hz-3\hz^2-\hz^3)
\over(\lambda+3\hz^2+2\hz^3)^2}\ .}
{}From the latter we see that ${\p T/\p\hz}=0$ at $\hz=\lambda^{1/3}$.
For $\lambda<1$, ${\p T/\p\hz}$ also has another zero in the physical
range $0\le\hz\le\lambda^{1/3}$. Hence for $\lambda\ge1$,
$T$ is a monotonic function increasing from 0 at $\hz=0$
to 
\eqn\Tmax{
T(\hz=\lambda^{1/3})={A\over2\pi}\lambda^{2/3}\ .}
For $\lambda<1$, $T$ rises from 0 at $\hz=0$
to a maximum at some intermediate value of $\hz$, and
then decreases down to a local minimum at $\hz=\lambda^{1/3}$
given by the same formula as in eq.~\Tmax.

In terms of $\hz$, the entropy  \btzentropy\ becomes
\eqn\btzSS{
S={2\pi\over G_4 A^2}{\hz\over\lambda+3\hz^2+2\hz^3}\ .}
Here again we also consider
\eqn\btzDS{
{\p S\over\p\hz}={2\pi\over G_4 A^2}{\lambda-3\hz^2-4\hz^3\over
(\lambda+3\hz^2+2\hz^3)^2}\ .}
The latter always has a zero in the
range $0\le\hz\le\lambda^{1/3}$. Hence the entropy increases
from zero at $\hz=0$ to a maximum at some intermediate value of $\hz$, and
then decreases down to 
\eqn\btzSend{
S={2\pi\over 3G_4 A^2}{1\over\lambda^{1/3}(1+\lambda^{1/3})}\ }
at $\hz=\lambda^{1/3}$, in agreement with \constarea.

Using the first law \firstb, we can now integrate
\eqn\firstc{
{\p  M_4\over \p\hz}=T{\p S\over\p\hz}}
with the boundary condition that $M_4=0$ at
$\hz=0$. A short calculation yields
\eqn\massfo{
M_4={1\over G_4 A}{\hz^2(1+\hz)
(\lambda-\hz^3)\over(\lambda+3\hz^2+2\hz^3)^2}\ .}
We want to compare this now to \mthreex. Expressing the latter in terms
of $\hz$, one finds 
\eqn\massth{
M_3={1\over 2G_3}{\hz^2(1+\hz)
(\lambda-\hz^3)\over(\lambda+3\hz^2+2\hz^3)^2}\ }
so that using \gnewt\ we get precise agreement $M_3=M_4$,
just as was found in \rEHM\ for flat branes.

The thermodynamical analysis for $k=-1$ 
can be carried through in close analogy to
the analysis for BTZ black holes above. As a matter of fact, if
we express the mass, entropy and temperature in terms of the variable
$\hz$ of \btzz\ we obtain precisely the same results \btzTT, \btzSS,
\massfo, and \massth, only this time we have $\hz^3>\lambda$. In fact,
we can also recover the results for $k=0$ by setting $\hz^3=\lambda$.
Hence, letting $\hz$ vary over $0\leq \hz< \infty$, these formulas give
us $M$, $S$ and $T$ for the full range of black holes with $\lambda>0$
and arbitrary $k$.

If we compute the thermodynamic
mass $M_4$ of the BTZ black
strings discussed above, we again find $M_3=M_4$.
To see this, notice that for a black
string, the four-dimensional entropy computed from the horizon area \btzstring\
agrees with the three-dimensional entropy computed from the circumference
\btzfirst:
\eqn\stringent{S={{\cal A}\over 4 G_4}={\pi\over A}\sqrt{2 M_3\over
\lambda G_3} ={{\cal C}\over 4 G_3}.} 
where we have used \gnewt\ and the fact that
the geometry on the brane is exactly \btzmet\ with
$\l3=1/A\sqrt{\lambda}$. The temperature on the brane must also agree with
the temperature in the bulk since it is constant over the horizon.
Since the temperature and entropy both agree, the mass obtained by
integrating the first law must also agree.

\subsec{Stability analysis}

In the range $0\le M_3 \le 1/24G_3$, there are two localized black hole
solutions as well as the black string. Which of these solutions 
gives the correct stable configuration for a given mass?
In \rCHR\ it was suggested that the Gregory-Laflamme instability \rGL\ 
should play an important role in determining the stable solution.
Gregory and Laflamme observed, in the context of Kaluza-Klein
compactification, that when a black string becomes longer than its
transverse size, unstable modes arise which appear to lead to the
black string breaking up into localized black holes. In \rCHR, this
result was combined with the observation that
in AdS space the only metric fluctuations that can
be supported have proper wavelengths shorter than the AdS scale $\ell$.
Hence it was argued that when the transverse size of a black string
reaches $\ell$, the Gregory-Laflamme instability should set in
and cause the black string to break up forming a localized black hole.
Their arguments were made in the context of discussing black strings and black
holes in the Randall-Sundrum scenario \rRS\ with an infinite transverse
dimension, but presumably they extend to the other scenario \rRSfirst\ with
a finite transverse dimension. This is the context in which we would
like to consider this mechanism here. Because of the negative curvature
on the branes, the minimum transverse size of the black string actually
arises at the `throat,' \ie $r=\sqrt{\la}\l4$, rather than at the second
brane. In any event, the fact that the black strings only
shrink to a minimum size (which is some fixed fraction
of the transverse size on the brane) provides a explanation as to
why black strings should appear as the only (stable) solution for
large masses. This result is in fact very similar to ordinary
Kaluza-Klein compactification where black strings prove to be
the appropriate solution for large masses. Essentially there, localized
black holes with large masses do not fit inside the compact
dimensions.

However, presumably the range $0\le M_3 \le 1/24G_3$ is a 
regime\foot{Recall this regime corresponds to the $k=+1$ black holes.} where
the black strings may break up and become localized. Rather than attempting
a detailed fluctuation analysis \rGL, we will argue stability of a given
solution on the grounds that it maximizes the entropy for a given mass.
For small $M_3$, the black string entropy \stringent\ is approximately equal to
that of black holes with small $mA$. In accord with the physical
picture of the horizon of these black holes found in section 3.2,
one can show that the black hole entropy is slightly smaller than
that of the black string. From Fig. 1 one sees that in fact
the black string entropy exceeds that for all of the black holes
in the range $0\le mA\le 2/\sqrt{27}$ --- recall that the upper
limit corresponds to the maximum mass \maxmthree. On the other
hand, for large $mA$, $M_3$ approaches zero again while the entropy
has the finite limit in eq.~\btzSend, in contrast
to the vanishing entropy of the black strings. Hence for small $M_3$, these
localized black holes will provide the stable configuration.
This is presumably a situation where the strings
are unstable and pinch-off to form a black hole. In order to
verify the picture put forward in ref.~\rCHR, one should determine
the minimum mass where the black string entropy still dominates
over that of the black holes. One could then verify that this
crossover occurs when the minimum transverse size of the black
strings reaches the AdS scale. 

\midinsert{ \centerline{\epsfysize2.5in\epsfbox{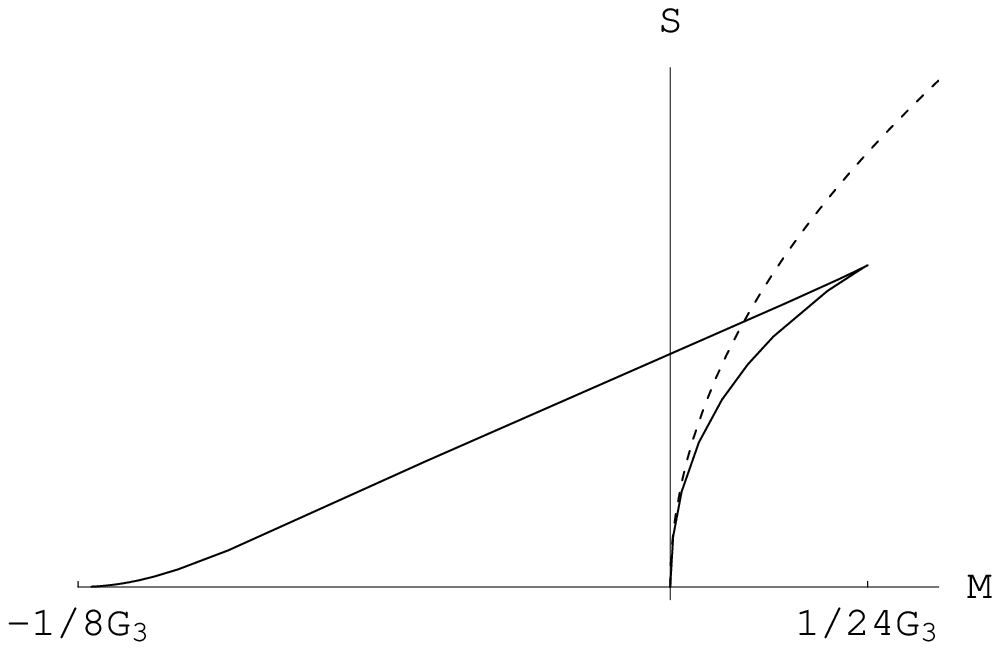}} 
Fig.~1: Entropy versus mass for BTZ black
strings (dashed line), BTZ-like black holes
(solid line, $M\geq 0$), and $k=-1$ localized black holes in
$AdS_3$ (solid line, $M<0$). The black
hole curve starts at $S=M=0$ for $mA=0$, and then grows to a maximum of
the mass and entropy for
$mA=2/\sqrt{27}$. As $mA\rightarrow\infty$ the solutions tend to
$M= 0$, $S$ finite (eq.\btzSend).} 
\endinsert

At the crossover point, the black string and black hole
masses, eqs.~\massstr\ and \massads\ (with $k=+1$), are identified
and hence $\Delta\vp$ is the same for both solutions. Also equating
the black hole and black string entropies, eqs.~\btzSS\ and
\stringent\ (substituting eq.~\massth\ in the latter), yields
a fourth order polynomial in terms of the auxiliary variable
\eqn\poly{\hz(\la-\hz^2-\hz^3)=0\ .}
Here, the root $\hz=0$ simply indicates that the mass and entropy
of the small black holes matches that of the black strings.
Denoting the solution of the crossover point as $\hz_x$,
it satisfies
\eqn\polyx{\hz_x^2+\hz_x^3=\la\ .}
While it would be possible to analytically determine the solution
of this equation, we did not find such an explicit solution
very illuminating. The circumference of the black string at the `throat' 
was calculated to be $C_{min}=\Delta\vp/A\sqrt{\la+1}$ in section 3.1.
Hence using eq.~\defell, we have
\eqn\ratio{
{C_{min}\over2\pi\l4}={\Delta\vp\over2\pi}={2\sqrt{\la}\hz_x\over
3\la+\hz_x^2}\ .}
Using eq.~\polyx, this result is implicitly a function of only $\la$.
Let us consider the result for small $\la$\ for which
$\hz_x^2\approx\la$ and hence eq.~\ratio\  yields
\eqn\ratios{
{C_{min}\over2\pi\l4}\approx{1\over2}\qquad{\rm for\ small}\ \la.}
So at the transition point, the minimum size of the black string horizon
is indeed of order the $AdS_4$ scale.
This result is certainly consistent with the instability mechanism
suggested in ref.~\rCHR. On the other hand, for large $\la$, one
has $\hz_x^3\approx\la$ and hence
\eqn\ratiol{
{C_{min}\over2\pi\l4}\approx{2\over3\la^{1/6}}\ll 1
\qquad{\rm for\ large}\ \la.}
Hence we seem to have found an inconsistency with the
discussion of ref.~\rCHR.

What has gone wrong? To resolve this apparent contradiction, we take a closer
look at the spacetime geometry. First recall
\eqn\recall{
\l4={1\over A\sqrt{\la+1}}\ ,
\qquad
L_1={1\over A}\ ,
\qquad
L_2={\sqrt{\la}\over A}\ .}
Now in the regime of small $\la$, we have
\eqn\lineaa{
L_1\approx\l4\gg L_2\ ,}
and hence the first brane at $r=L_1$ plays the role of the Planck brane,
\ie it inherits the larger scale factor from the AdS geometry.
Thus our construction and analysis are consistent with the suggested
instability mechanism \rCHR. That is, as the mass of the black string
shrinks (by, \eg, Hawking radiation) eventually its minimum
cross-section reaches the AdS scale $\l4$. At this point, the
Gregory-Laflamme instability causes the string to break up and
form a black hole localized in the vicinity of the Planck brane
at $r=L_1$. On the other hand, in the regime of large $\la$, we have
\eqn\linebb{
L_2\gg L_1\gg\l4\ ,}
and hence the second brane at $r=L_2$ plays the role of the Planck
brane. Hence using the mechanism of \rCHR, we would argue that as the
mass of the black string shrinks, it would become unstable
when $C_{min}/2\pi\l4\approx1$ and the string would then break
up to form a black hole localized at the second brane, {\it not}
the brane at $r=L_1$. Hence while the result in eq.~\ratiol\ is
not incorrect, it is probably not relevant to determining the
stable configuration in this situation. That is we are comparing
the black string to the `wrong' black hole solution. Instead we expect
the true stable solution to be a black hole localized on the brane
at $r=L_2$, but our construction in section 3 has not provided
us with this solution.

In short, in the regieme where we can reliably compute the transition
between the black string and black hole, our results are consistent with
the estimates in \rCHR.

\newsec{Adding rotation}

It is easy to include the effects of rotation into the picture. What we
need is a metric that describes an accelerating, rotating black hole in
a spacetime with a negative cosmological constant. Such a solution can
be found as a member of a subfamily of solutions of the most general
type D metric\foot{A type D metric is one in which the Weyl tensor is
algebraically special and has two pairs of repeated principal null
directions \rWald. The ``D" stands for ``degenerate" and has nothing to do with
D-branes or D-terms.} constructed in \rPD\ (the complete family of
neutral solutions possesses, in addition, a NUT parameter, which we have
set to zero here). The solution we are interested in is
\eqn\adsd{\eqalign{ds^2=&{1\over A^2 (x-y)^2} \biggl[ {H(y)\over
1+a^2x^2y^2}(dt+ax^2d\varphi)^2 -{1+a^2x^2y^2\over H(y)} dy^2
\cr
&+{1+a^2x^2y^2\over G(x)}dx^2 +{G(x)\over
1+a^2x^2y^2} (d\varphi-ay^2dt)^2 \biggr],\cr}}
where
\eqn\hy{H(y)=-\lambda +ky^2 -2 m Ay^3-a^2 y^4}
\eqn\gx{G(x)=1+ kx^2 -2m Ax^3+a^2\lambda x^4}
and $k=+1,0,-1$. Again, the metric satisfies $R_{\ssc AB}=-(3/\l4^2)
g_{\ssc AB}$,  with $\ell_4$ defined as in \defell. The AdS
C-metric \cads\ is recovered by simply setting $a=0$.

When $m=0$ the geometry is once again that of AdS$_4$ in disguise.
In
this case, the change of coordinates that makes this manifest is
\eqn\changeD{r={\sqrt{y^2+\lambda x^2}\over
A(x-y)},\qquad
\rho=\sqrt{1+kx^2-a^2 x^2 y^2\over y^2+\lambda x^2}.}
Then \adsd\ becomes
\eqn\niceD{ds^2={dr^2\over{r^2\over \l4^2}-\lambda}+r^2\left[ -\left(
\lambda\rho^2-k+{a^2\over \rho^2}\right)dt^2+{d\rho^2\over 
\lambda\rho^2-k+{a^2\over \rho^2}}+\rho^2\left(d\varphi-{a\over
\rho^2}dt\right)^2\right].}
The three-dimensional metrics in brackets have constant Riemann
curvature. Notice that, as before, these sections at constant $r$ are
sections at constant $x/y$. Now, however, the coordinate systems
are rotating. For $\lambda>0$ and $k=+1,0$, these sections are
precisely the geometries of rotating BTZ black holes. So the entire spacetime
can be viewed as a rotating BTZ black string.

The metrics \adsd\ present features familiar from the Kerr
metric. For example, when $m\neq 0$ they have a curvature
singularity at
\eqn\ring{{1\over y^2}+a^2 x^2=0,}
which is closely analogous to the ring singularity of the Kerr solution
at $r^2 + a^2\cos^2\theta =0$ in standard Boyer-Lindquist coordinates.
On the other hand, at the zeroes $x_i$ of $G(x)$ the Killing vector
\eqn\axial{{\p\over\p\varphi} -ax_i^2 {\p\over\p t}} has a fixed-point
set. These are the rotation axes. In order to avoid a conical defect
at one of them, say at $x=x_2$, then we have to identify points along
the integral curves of \axial\ with the appropriate period. This corresponds
to changing coordinates from $t,\vp$ to $\tilde t,\vp$ where
\eqn\deftild{\tilde t = t + a x_2^2 \vp}
and choosing the period of $\varphi$ to be given by \delphi,
on surfaces of fixed $\tilde t$.\foot{Of course, periodic identification
of $\vp$ at constant $\tilde t$ is not the same as periodic identification of
$\vp$ at constant $t$.}

In a similar way, at the roots $y_i$ of $H(y)$ the Killing vector
\eqn\horiz{{\p\over\p t} + ay_i^2 {\p\over\p\varphi}} becomes null. These
correspond to horizons with  angular velocity $\Omega=ay_i^2$. Actually,
this is appropriate only if we are considering black strings with no
rotation axis. As we have just seen, if there is an axis, regularity requires
that we use $\tilde t,\vp$ coordinates. In this case, the angular velocity
is 
\eqn\angvel{\tilde\Omega={ay_i^2\over
1+a^2x_2^2 y_i^2}}
 Observe that when
$a\neq 0$, $H(y)$ will typically have one more root than in the case
$a=0$. This corresponds to the addition of an inner horizon which
is characteristic of rotating black holes.  

The construction of a two-brane can be performed, as before, at $x=0$ (and
also at $y=0$, which will be needed when $\lambda>0$). To analyze the
metric on the brane, let
$\rho=-1/y$, so the
section at $x=0$ is
\eqn\branegeo{ds^2={1\over A^2}\biggl[-\left(\lambda\rho^2-k-{2mA
\over \rho}+{a^2\over \rho^2} \right)dt^2+{d\rho^2\over
\lambda\rho^2-k-{2m A
\over \rho} +{a^2\over \rho^2}}+\rho^2
\left(d\varphi-{a\over
\rho^2} d t\right)^2\biggr].}
When $\lambda>0$ and $k=1$, this is 
similar to the rotating BTZ metric, only with
additional $2 mA/\rho$ terms  due, as we discussed earlier, to the 
four-dimensional origin of the black hole. The discussion following \wallgeo\
applies to this situation as well. 
In the following we focus on the cases
of asymptotically flat branes, $\lambda=0$, and asymptotically $AdS_3$
branes, $\lambda>0$. 

\subsec{$\lambda=0$: Kerr black holes on the brane}

Black holes on asymptotically flat branes were studied, in the absence
of rotation, in \rEHM --- the notation used there corresponds
to $\mu=mA$ and $\ell=1/A$. Here we will focus only on the new features
introduced by a non-vanishing $a$. 

When $\lambda=0$ ($A=1/\ell_4$) we have to set $k=-1$, in order that
$\p_t$ is timelike at infinity on the brane. When $m=0$, the roots of 
$G(x)$ are just $x_i =\pm 1$, and $|G'(x_i)|=2$
so to avoid conical singularities on
these axes we need to set $\tilde t = t+ a\vp$
as in eq.~\deftild\ and identify $\vp$
with period $2\pi$ \delphi. The geometry
is that of $AdS_4$, but instead of the coordinates
\changeD\ it is more convenient to define
\eqn\tohors{z=1- {x \over y} ,
\qquad \tilde\rho = -{\sqrt{(1-x^2)(1+a^2y^2)}\over y}}
which brings the metric \adsd\
into Poincare coordinates
\eqn\hors{ds^2={\ell_4^2\over z^2 }(-d\tilde t^2 +d\tilde\rho^2
+\tilde\rho^2 d\varphi^2+dz^2 ).}
Apart from $z$ being essentially the inverse of $r$, the choice of
radial coordinate on the brane is slightly different from
\changeD\ with $\lambda=0$. 

Now consider the case $m>0$. 
As was the case in the absence of rotation, when $\lambda=0$ the
surface
$y=0$ is a degenerate cosmological horizon analogous to $z=\infty$ in
\hors. When $|a| < mA$ there are two other zeroes of $H(y)$, at
$y=y_{0\pm}=-{mA\over a^2}\pm {\sqrt{m^2A^2-a^2}\over a^2}$, which are
interpreted as inner and outer black hole horizons. In the extremal limit
$|a|=mA$, they coincide at $y=-1/mA$. If $|a|>mA$ the singularity would
be naked.

The metric on the brane at $x=0$
is locally equivalent to the equatorial slice of
the four-dimensional Kerr metric. To see this, set $\hat t = (t+a\vp)/A$,
$\hat \rho = \rho/A$. Then since $\la =0$ and $k=-1$, \branegeo\ becomes
\eqn\branegeom{\eqalign{ds^2=&-\left(1-{2m
\over \hat\rho} \right)d
\hat t^2+ \left(1-{2m \over \hat \rho} 
+{a^2\over A^2 \hat\rho^2}\right)^{-1}
d\hat \rho^2 \cr
&+\left(\hat \rho^2+{a^2\over A^2} +
{2ma^2\over A^2 \hat\rho}\right)d\varphi^2-{4a m\over A\hat \rho}
d\hat t d\varphi.\cr}}
This  is indeed the equatorial section of the
Kerr solution, with  mass $G_4 M_4=m$
and rotation parameter $a/A$.

Even though the metric on the brane
is locally equivalent to Kerr it is not globally
equivalent. This is because the identification we need to remove the
conical singularities on the axis forces
the solution to rotate at infinity. This can be seen as follows.
Observe that when $\lambda=0$ the function $G(x)$ is the same as in
the
absence of rotation. This means that the 
axis of $\varphi$ lies along the same value
of $x_2$ as the nonrotating AdS C-metric, and
the periodicity \delphi\ is determined only in
terms
of $mA$ and not of $a$. However \deftild\ shows that we must
hold $\tilde t = t+a x_2^2 \vp$ (rather than $t+ a\vp$)
fixed when we identify $\vp$. If we set
$t = \tilde t - a x_2^2 \vp$ into \branegeo\ (with $\la=0$ and $k=-1$),
then to leading order the metric is
\eqn\asymprot{ds^2 = {1\over A^2}[ -d\tilde t^2 - 2a(1-x_2^2) d\tilde t d\vp +
           d\rho^2 + \rho^2 d \vp^2]}
We now want to compare this with the
 metric that describes a particle of spin $J_3$ in $2+1$
dimensions \rthreeo\
\eqn\spinp{ds^2=-(dt+4
G_3 J_3d\varphi)^2+d\rho^2+\rho^2 d\varphi^2}
(with $\vp$ periodically identified on fixed $t$ surfaces).
Clearly 
\eqn\comj{ J_3 = {a(1-x_2^2)\over 4AG_3}}

For small $mA$, $x_2 \approx 1-mA$, and we have seen that the four-dimensional
mass and angular momentum of the Kerr black hole are $m = G_4 M_4$ and
$J_4 = M_4 a/A$.
Substituting into \comj\ and using \gnewt\ to relate the Newton's 
constants, we find
\eqn\equalj{ J_3 =J_4}
as well as $M_3=M_4$.
So even though the angular momentum in $2+1$ gravity is a global effect,
it turns out to agree exactly with the four-dimensional angular momentum
of the black hole. This is directly analogous
to what we found for the mass in the absence of rotation \rEHM, and
can be understood by a similar argument.
Before introducing the
brane, the AdS C-metric describes a black hole accelerating in AdS. 
The cause of the acceleration is a cosmic string pulling on the black hole.
When the string is attached to a spinning horizon, it is set into rotation.
This rotation, like the conical deficit angle, is a global effect that can be
detected on every section transverse to the axis. When the brane is
introduced, the cosmic string is removed, but the boundary conditions on 
the brane essentially reproduce its effect. We expect that the agreement
\equalj\ continues to hold for large $mA$.

\subsec{$\lambda >0$: Rotating BTZ black holes}

When $\la >0$, the metrics on the brane \branegeo\
are related to those of rotating
solutions in $AdS_3$. In fact, if
$mA=0$ they are the same as BTZ
black holes for $k=+1,0$,
and negative mass particles for $k=-1$.  Thus, the mass and spin
on the brane can be measured by comparing to the spinning BTZ solution
\rBTZ,
\eqn\kerrbtzmet{\eqalign{
ds^2=&-\left({\hat \rho^2\over\l3^2}-8G_3M_3+{(4G_3J_3)^2\over 
\hat\rho^2}\right)d\hat t^2+
\left({\hat \rho^2\over\l3^2}-8G_3M_3+{(4G_3J_3)^2\over 
\hat\rho^2}\right)^{-1}
d\hat\rho^2 \cr
 &+\hat\rho^2\left(d\hat\vp-{4G_3J_3\over \hat\rho^2}d\hat t\right)^2\cr}}
where $\hat\vp$ has periodicity $2\pi$ (for fixed $\hat t$).

The most important feature introduced by a nonzero value of the
rotation parameter $a$ appears in the structure of $G(x)$. Notice that
since $G(x)$ is now a quartic polynomial we are not guaranteed to always
have a real root, even when $m\neq 0$. Consider for instance
starting from a rotating BTZ black string ($k=+1$, $m=0$), which extends
throughout the entire spacetime. Now turn on a small, but nonzero $m$.
When $a=0$ we have seen that this results in the string being chopped
off at one of its ends, the endpoint being marked by the smallest
positive real root of $G(x)$. However, when $a\neq 0$ there is a range
of (small) values of $mA$ such that all roots of $G(x)$ are still
complex, so the range of $x$ is unrestricted, $-\infty<x<\infty$ (it
becomes $0\leq x<\infty$ once the brane is introduced). Therefore, for
$mA$ within this range of values, we still find a black string, which,
runs from the brane at $x=0$ to the brane at $y=0$. 

As we let $mA$ grow larger we reach a certain critical value at which
$G(x)$ develops a positive double root $x_2$. Then the proper
spatial distance to $x_2$, given as $\sim \int^{x_2} dx/\sqrt{G(x)}$,
is infinite, so the black string has infinite proper length now\foot{The
area remains finite, though.}. The string extends infinitely far even if
we put a second brane at $y=0$ to compactify spacetime. The way this can
happen is that this second brane also develops an infinite funnel
centered at the double root $x=x_2$. The black string then extends down
this funnel. Notice the period $\Delta\varphi$ can still be arbitrarily
chosen.

For larger values of $mA$, the function $G(x)$ has real positive simple
zeroes, and the upper value for $x$ is the smallest of these zeroes,
which is now a finite proper distance away. This marks the outermost
extent of the black hole horizon inside the bulk. It is only in this
case that the period $\Delta\varphi$, and hence the mass on the brane,
is fixed (to the value \delphi) once $mA$ and $a$ are fixed. Let us note
that whether the horizon is extremal or not does not appear to influence
the discussion of these issues.

These `localized' black holes
also produce a three-dimensional angular momentum, just as in the
asymptotically flat case, $\la=0$. The total angular momentum
detected on the brane is obtained by comparing the asymptotic geometry to
\kerrbtzmet. As we saw earlier, a nonsingular rotation axis
requires that the angular coordinate be identified periodically on
surfaces of constant $\tilde t$ \deftild. 
So we change the time coordinate in
\branegeo\ by $t=\tilde t-a x_2^2\vp$
so that points now are identified as $(\tilde t,\rho,\vp)\sim(\tilde
t,\rho,\vp+\Delta\vp)$.
It is easy to see that this results in the coefficient of the
$d\tilde td\vp$ growing like $\rho^2$, instead of a constant, as in
\kerrbtzmet, which might appear to imply a
diverging angular momentum. This, however, is an artifact of
the coordinate system, and not a true global effect. To see this,
notice that we can still define $\tilde \vp=\vp + f(\tilde t,\vp)$ without
changing the global identifications.
In particular, letting $\vp=\tilde
\vp-{\la a
x_2^2\over 1-\la
a^2x_2^4}\tilde t$ we can remove the $\rho^2$ term in $d\tilde td\tilde \vp$,
and finally find
the asymptotic metric
\eqn\rotasym{A^2ds^2=-{\lambda\over 1-\lambda
a^2x_2^4}\rho^2d\tilde t^2+\rho^2(1-\lambda a^2
x_2^4)d{\tilde \vp}^2+
{4 a mA x_2^3\over 1-\la a^2x_2^4}d\tilde t
d\tilde \vp +{d\rho^2\over \lambda\rho^2}+\dots}
The spin is then
\eqn\spinbtz{J_3=-{a m x_2^3\over 2
G_3}\left({\Delta\vp\over
2\pi}\right)^2.}
For a small range of parameters (corresponding to $x_2$ close 
to becoming a double root) $\tilde t$ becomes spacelike and $\tilde \vp$
becomes timelike in \rotasym. Since $\tilde \vp$ is periodic, these
solutions have closed timelike curves near infinity and are presumably
unphysical.

\newsec{Discussion}

So far we have been considering black holes on branes with
asymptotically flat or AdS geometries. It is easy to extend the
discussion to the case of $\lambda<0$, where the branes have compact
volume and the geometry of deSitter universes. The analysis of the AdS
C-metric in this case is quite straightforward, since apart from the
different asymptotics at four-dimensional infinity, it is quite similar
to the C-metric without a cosmological constant (which corresponds to
$\lambda=-1$). Setting $k=-1$,  it
follows that for small $mA$, $H(y)$ has two negative roots\foot{For
simplicity we only discuss the non-rotating case.}. The smaller root
corresponds to a black hole horizon of finite size and the larger one is
an acceleration horizon. The solution can be continued past the
acceleration horizon, to find a second black hole accelerating in the
opposite direction. When $mA$ grows to a value such that $H(y)$ develops
a double root, the black hole horizon and the acceleration horizon
coincide. Limiting situations of this sort have been considered in \rHS.
For larger values of $mA$ the singularity at $y=-\infty$ is naked. 

On the brane, for small $mA$ we find a black hole inside a $dS_3$
universe. Since there are no black holes in $2+1$ gravity with a
positive cosmological constant, these are entirely a result of the
modifications coming from the extra dimension. If the metric is
continued past the three-dimensional cosmological horizon, then we
recover a second black hole, which corresponds to the second black hole
in AdS$_4$. The appearance of the naked singularity as $mA$ grows
corresponds to the black hole horizon on the brane becoming coincident
with, and then larger than the cosmological horizon. We are not aware of
any satisfactory definition of mass in $dS_3$, but we might expect (or
define) the mass of black holes on the brane to be equal to the mass
$M_4$ computed from thermodynamics in the bulk.

We have seen in section 4.2 that the solutions we have constructed do
not appear to be the most general ones describing black holes on
two-branes. In light of this, it is natural to wonder if the existence
of a maximum possible mass of order the Planck scale for a localized
black hole on the brane (discussed in section 3.2) is just an artifact
of our family of solutions. While we do not know how to construct the
most general solution, we now comment on possible alternative locations
of the branes in the AdS C-metric. In contrast to the situation where no
black holes are present (and the branes can be placed at any surface at
constant $x/y$), when $m>0$ the condition for the location of the branes --
that the extrinsic curvature be proportional to the induced metric -- is
highly restrictive. Above, we found it was possible to place the
two-branes at $x=0$ and $y=0$. More generally, for the metrics $\cads$,
the equation $K_{\mu\nu}\propto g_{\mu\nu}$ is solved on any surface
$x=x_b$ such that $G'(x_b)=0$ and $G(x_b)\neq 0$. Similarly, it is solved
on any surface $y=y_b$ with $H'(y_b) =0$ and $H(y_b) \ne 0$.
However it turns out that the resulting
solutions are equivalent to the ones already
studied.

Let us consider the case of constant nonzero $x$ in more detail.
For $k=\pm
1$ it is possible to place a brane at the
second root of $G'(x)$, $x_b= k/3mA$. However,
one can always shift coordinates so that the brane is at $x=0$ in the
new coordinates. Define $\gamma=\sqrt{G(x_b)}=\sqrt{1+{k\over 27 m^2
A^2}}$, and change
\eqn\map{\eqalign{
x= x_b +\gamma \bar x,&\qquad y=
x_b+\gamma \bar y,\cr
t= {\bar t\over\gamma},&\qquad \varphi=
{\bar\varphi\over\gamma}. \cr}}
To convert the resulting metric into one of the same form as \cads\ we
just have to redefine the parameters as
\eqn\mappar{A=
{\bar A\over\gamma},\qquad k= -\bar k ,}
while leaving $m$ unchanged. Notice $x=x_b$ is mapped onto $\bar x=0$.
One can
also see that the sign of $\bar\lambda=- 1+1/l^2\bar A^2$ may be
different to that of $\lambda$. 
For illustration, consider placing a two-brane at $x_b=-1/3mA$ when
$\lambda=0$ and $k=-1$. In order for $x_b$ to fall into the range of
variation of $x$, we must require $mA>1/3\sqrt{3}$ (this also ensures
$\gamma$ is real) which is in a black string regime. 
The transformation above results into $\bar k=+1$ and
$\bar\lambda=1/(27 m^2A^2-1) >0$. Conversely, if $k=+1$ and
$\lambda=1/27m^2A^2$, then a two-brane at $x=1/3mA$ is asymptotically
flat. 

From our investigations in sections 3 and 4,
we arrive at the following general picture:
For $M_3>1/24G_3$, a gravitationally collapsed system can only form
a BTZ black string. These black strings remain the stable
configuration down to a certain transition mass, $M_3\approx1/32G_3$ 
for small $\lambda$, where the minimum transverse size of the black
string shrinks to be of order the four-dimensional AdS radius.
(Recall that our calculations of this transition are reliable only for 
$\la <1$.)
At around this mass, the black strings are destabilized
by the Gregory-Laflamme instability \rGL, which causes them to break up
and form a black hole localized on the Planck brane. In the small
$\la$ regime, one finds that at the transition point the localized
black hole appears to have essentially the geometry of a large
($\rho_{\rm horz}\approx\la^{-1/2}$) BTZ black hole. That is, the
$2mA/\rho$ corrections in eq.~\wallgeo\  are very small everywhere
outside of the event horizon. In addition, at the transition point
the horizon of these localized black holes extends out in the
transverse space to come very close to the `throat' at $r=\sqrt{\la}
\l4$.  Below the transition point down to $M_3=0$, 
black holes localized on the Planck brane are the stable end-point
of a gravitational collapse. This continues to be true for $-1/8G_3<M_3
<0$, where these black holes are the only solutions with an event horizon.
As $M_3$ approaches $-1/8G_3$, these black holes resemble 
four-dimensional Schwarzschild AdS black holes. At precisely $M_3=-1/8G_3$,
one is left with a slice of pure AdS$_4$ in the bulk spacetime
on either side of the branes.

By considering the instability of the black string,
one is led to 
a general argument for the sign of the correction terms to the gravitational
potential on the brane. The following argument applies in all dimensions,
but to be definite, we consider the standard case of a three brane in $AdS_5$.
The usual four dimensional Schwarzschild black hole can be realized on the
brane if the full solution is a black string \rCHR. 
When the black string is unstable, it caps off its ends to form a localized 
black hole. But this will always decrease the total horizon area unless the
cross sectional area is increased at the same time.
So  if a localized black hole is going to be stable,
it must have {\it larger}
horizon area on the brane than the original Schwarzschild
solution of the same mass.
This tells us the sign of the corrections to the gravitational
potential. That is, if we have
\eqn\potter{
g_{tt}=-1+V(r)=-1+{2 GM\over r}+\alpha \left({L\over r}\right)^\beta
{2 GM\over r}}
then we had better have $\alpha>0$ in order that the horizon be
at a larger radius than $r_H=2 GM$. This agrees with the sign obtained
from a perturbative calculation in \rRS\ (see also \rGT, \rGKR), and from
our exact solutions in one lower dimension in section 3. 

In light of the AdS/CFT correspondence, it has been suggested that the
Randall-Sundrum scenario can be viewed as a coupling of the lower
dimensional gravity on the brane to a strongly coupled conformal field
theory which is dual to gravity in the bulk \rGubser. For the case of negative
curvature branes, there is the intriguing possibility that we could
apply this duality twice. Since the theory on the brane is 2+1 AdS
gravity (coupled to a $2+1$ CFT), it may be equivalent in some sense to
a $1+1$ CFT! In a sense, one may have a holographic description of a
holographic description of the theory.

\vskip 2cm

\bigbreak\bigskip\bigskip\centerline{{\bf Acknowledgements}}\nobreak

\vskip .5cm

GTH and RCM would like to thank the participants of the ITP Program on
Supersymmetric Gauge Dynamics and String Theory for stimulating
discussions, especially Steve Giddings, Lisa Randall and Raman Sundrum.
The work of RE is partially supported by UPV grant 063.310-EB187/98 and
CICYT AEN99-0315. GTH was supported in part by NSF Grant PHY95-07065.
RCM was supported in part by NSERC of Canada and Fonds du Qu\'ebec. At
the ITP, RCM was supported by PHY94-07194. This paper has report numbers
EHU-FT/9915, McGill/99-40, and NSF-ITP-99-151.

\listrefs

\end